\documentclass[prl,twocolumn,superscriptaddress,showpacs]{revtex4-1}
\usepackage{amsmath,graphicx}
\usepackage[normalem]{ulem}
\usepackage{color,soul}
\usepackage{bm}

\begin{document}

\title{Exotic attractors of the non-equilibrium Rabi-Hubbard model}
\author{M. Schir\'o}
\affiliation{Institut de Physique Th\'{e}orique, Universit\'{e} Paris Saclay, CNRS, CEA, F-91191 Gif-sur-Yvette, France}
\author{C. Joshi}
\affiliation{Department of Physics, University of York, Heslington, York, YO10 5DD, UK}
\affiliation{SUPA, School of Physics and Astronomy, University of St Andrews, St Andrews KY16 9SS, UK}
\author{M. Bordyuh}
\affiliation{Department of Electrical Engineering, Princeton University, Princeton, New Jersey 08544, USA}
\author{R. Fazio}
\affiliation{ICTP, Strada Costiera 11, I-34151 Trieste, Italy}
\affiliation{NEST, Scuola Normale Superiore and Istituto Nanoscienze-CNR, I-56127 Pisa, Italy}
\author{J. Keeling}
\affiliation{SUPA, School of Physics and Astronomy, University of St Andrews, St Andrews KY16 9SS, UK}
\author{H. E. T\"ureci}
\affiliation{Department of Electrical Engineering, Princeton University, Princeton, New Jersey 08544, USA}
\date{\today}

\begin{abstract}
  We explore the phase diagram of the dissipative Rabi-Hubbard model,
  as could be realized by a Raman-pumping scheme applied to a coupled
  cavity array. There exist various exotic attractors, including
  ferroelectric, antiferroelectric, and incommensurate fixed points,
  as well as regions of persistent oscillations.  Many of these
  features can be understood analytically by truncating to the two
  lowest lying states of the Rabi model on each site.  We also show
  that these features survive beyond mean-field, using Matrix Product Operator
  simulations.
\end{abstract}

\maketitle

\emph{Introduction - }  A number of recent experimental breakthroughs \cite{kasprzak_boseeinstein_2006,baumann_dicke_2010,klaers_bose-einstein_2010,baden_realization_2014} have spurred the investigation of non-equilibrium properties of hybrid quantum many-body systems of interacting matter and light. Characterized by excitations with a finite lifetime, when sustained by finite-amplitude optical drives they display steady-state phases that are generally far richer \cite{bhaseen_dynamics_2012,keeling_collective_2010,carusotto_quantum_2013,ChangEtAlNatPhys06,
OtterbachEtAl_PRL13,HoningEtAlPRA13} than their equilibrium counterparts \cite{dalfovo_theory_1999,garraway_dicke_2011}. Critical phenomena in these open driven-dissipative systems often come with genuinely new properties and novel dynamic universality classes, even when an effective temperature can be identified~\cite{nagy_critical_2011,oztop_excitations_2012,torre_keldysh_2013,kulkarni_cavity-mediated_2013,brennecke_real-time_2013}, a statement that can be made robust in renormalization group calculations~\cite{sieberer_nonequilibrium_2014,SiebererReviewKeldysh_arxiv15}. 
Coupled cavity QED systems~\cite{Angelakis2007a,Hartmann2006,Greentree2006} have emerged as natural platforms to study many-body physics of open quantum systems. The current fabrication and control capabilities in solid-state quantum optics allows to probe lattice systems \cite{houck_on-chip_2012,loo_photon-mediated_2013,abbarchi_macroscopic_2013,eichler_quantum-limited_2014,jacqmin_direct_2014,raftery_observation_2014,mlynek_observation_2014,LeHurEtAl_arxiv15,BabouxEtAl_arxiv15}  in the mesoscopic regime, providing a first glimpse into how macroscopic quantum behavior may arise far from equilibrium.
It is therefore of interest to (i) identify a physical system where a non-equilibrium phase transition can be studied -- at least in principle -- in the thermodynamic limit, (ii) can be compared to an equilibrium analogue through a proper limiting procedure, and (iii) can be easily realized in an architecture that is currently available.

\begin{figure}[htpb]
  \centering
  \includegraphics[width=3.2in]{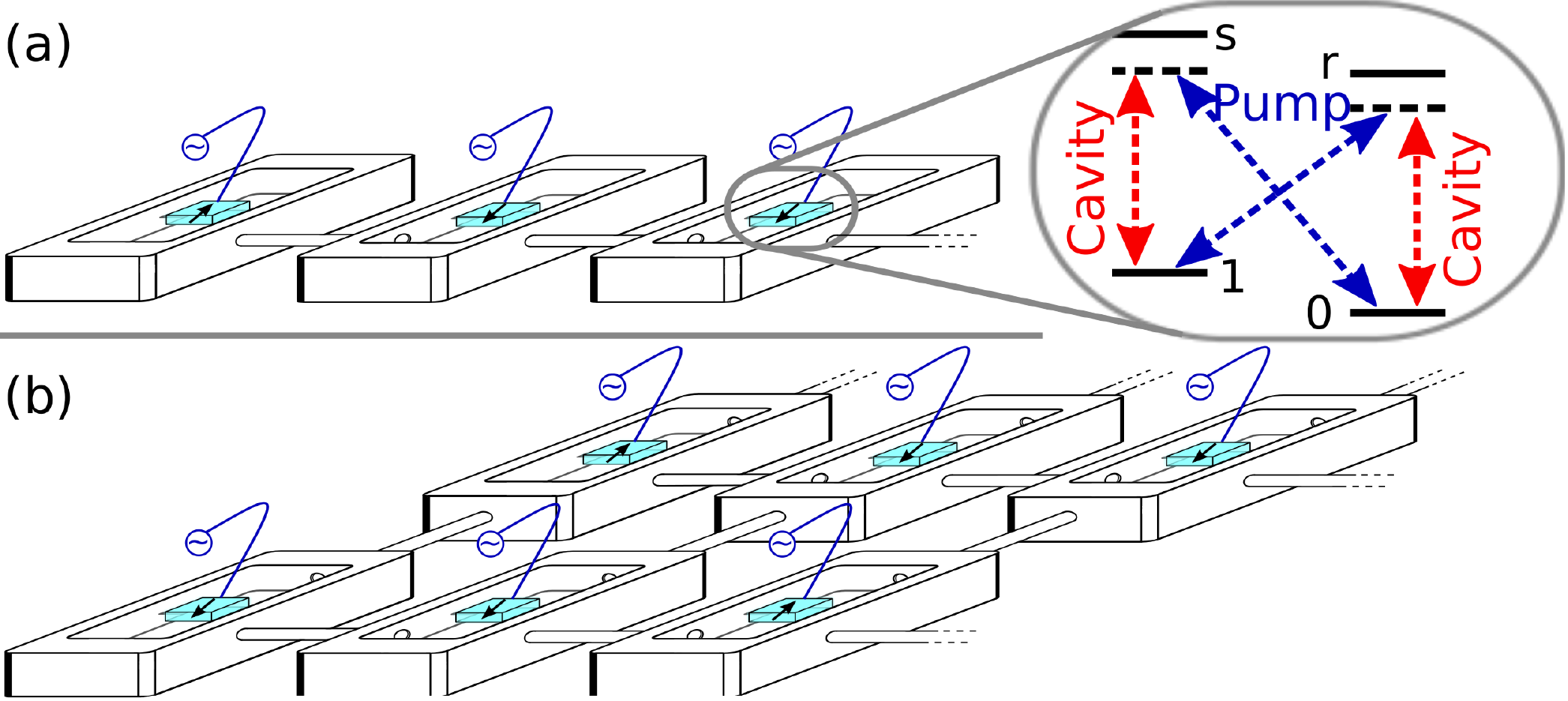}
  \caption{Schematic array of coupled cavities in (a) 1D or (b) 2D, containing "Raman-driven" qubits.  Inset: Cartoon of Raman
    driving: two low-lying levels of each artificial atom are
    connected via excited states.  The strength of the  drive
    determines the effective atom-cavity coupling~\cite{Note1}. }
  \label{fig:model}
\end{figure}
The Rabi-Hubbard (RH) model~\cite{schiro_phase_2012} represents the minimal description of coupled Cavity QED systems,  explicitly containing terms which do not conserve particle number. These terms are relevant for the low-frequency behavior of the coupled system and their inclusion lead, in equilibrium, to a $Z_2$-symmetry breaking phase transition between a quantum disordered para-electric phase and an Ising ferroelectric~\cite{schiro_phase_2012,schiro_quantum_2013}. 
The equilibrium RH transition requires a sizable inter-cavity hopping or light matter interaction, of the order of the transition frequency of cavities and qubits~\cite{schiro_phase_2012}. While such ultra-strong coupling regimes have recently been realized in specific circuit QED architectures~\cite{niemczyk_circuit_2010}, they are hard to achieve
in lattice Cavity QED settings. To overcome this challenge it is therefore  crucial to engineer effective realizations of the RH model by, e.g., suitably designed driving schemes. In this paper we study the behavior of such a scheme that leads to a RH model with highly tunable parameters and in a fully non-equilibrium regime. 

The interplay of drive and dissipation results in exotic attractors, remarkably different from thermal equilibrium, with rich patterns of symmetry breaking including incommensurate and antiferroelectric ordering.  In the following we  identify and explain these orders, and the associated phase transitions  using a variety of mean field approaches and then confirm the qualitative picture by simulating a one-dimensional open RH model with a Matrix Product Operator (MPO) approach~\cite{VidalPRL03,VidalPRL04,ZwolakVidalPRL04,OrusVidalPRB08,SchollwockAnnPhys11}.

\emph{Tunable Open Rabi Hubbard Model - } Recently several proposals to engineer effective light-matter interactions by suitable designed driving schemes have appeared~\cite{dimer_proposed_2007,Ballester2012,Grimsmo2013a,Zou2014}, based on a variety of platforms including superconducting circuit QED, impurities in diamond, and ultracold-atoms~\cite{baden_realization_2014}. Here for concreteness we consider a lattice of coupled cavity-QED systems, where on each lattice site $n$ there is a four-level system  which is driven and coupled to a cavity mode (see Fig.~\ref{fig:model}). We can write the full Hamiltonian as $\hat H=J \sum_{\langle nm\rangle} \hat a^\dagger_n \hat a^{}_m +\sum_n \hat h_n^{4LS}$ where $J$ is the hopping rate, $\hat a^\dagger_n, \hat a^{}_n$ are creation/annihilation operators of the cavity mode while $\hat h_n^{4LS}=\omega_0a^\dagger_n \hat a^{}_n+\sum_{j=0,1,r,s} E_j \vert j\rangle_n\langle j\vert_n+\hat H_{\rm int}+\hat H_{\rm drive}(t)$  describes the driven four-level atom coupled to cavity. The key idea of this Raman pumping scheme~\cite{dimer_proposed_2007} is that the cavity mediates transitions between states $0\leftrightarrow r,1\leftrightarrow s$ (blue arrows in Fig~\ref{fig:model}), i.e. $\hat H_{\rm int}= \hat a_n \left(g_r\vert r\rangle\langle 0\vert +g_s \vert s\rangle\langle 1\vert\right)+ h.c.$ while a two-frequency pump drives the transitions $1\leftrightarrow r,0\leftrightarrow s$ (red arrows in Fig~\ref{fig:model}), i.e. $\hat{H}_{\rm drive}(t)= \frac{\Omega_r}{2}\,e^{-i\omega_r^p\,t}\,\vert\,r\rangle\langle\,1\vert +
\frac{\Omega_s}{2}\,e^{-i\omega_s^p\,t}\,\vert\,s\rangle\langle\,0\vert+\text{H.c.}$.
The combined effect of light-matter interaction and drive is to induce an effective direct coupling between the two low lying atomic states. More formally, this can be shown by the standard procedure of first eliminating the explicit time-dependence of $H_{\text{drive}}(t)$ moving to a rotating frame and then eliminating the excited states to obtain an effective model for the cavity photon and the states $\vert 0\rangle,\vert 1\rangle$ acting as an effective qubit \footnote{See supplemental material at XXXXX for further details of the Raman driving scheme, the effective model, and further MPO results}.  The effective Hamiltonian takes the generalized RH form, 
\begin{align} \label{eq:1}
\hat H_{\rm RH}&=-J\sum_{\langle nm\rangle} \hat a^\dagger_n \hat a^{}_m+\sum_n h_n\\
\hat h_n &= \frac{\omega_0}{2} \hat \sigma_n^z + \omega \hat a^\dagger_n \hat a^{}_n+(g\hat a^\dagger_n \hat \sigma^-_n+g'\hat a^\dagger_n \hat \sigma^+_n+ hc)
\end{align}
where on each site $n$ we have now a two-level system, $
\hat \sigma^+=\vert 1\rangle\langle 0\vert$, $\hat \sigma^-=\vert 0\rangle\langle 1\vert$. The co-rotating and counter-rotating couplings $g,g^\prime$, as well as the effective cavity and qubit frequencies $\omega, \omega_0$  are tunable through the amplitude and frequency of the Raman drive~\cite{Note1}.  We stress that although described by a static Hamiltonian the problem retains its non-equilibrium character since cavity and qubit excitations are coupled to baths which are described, in the rotating frame,  by a non-thermal distribution of modes. To account for the dissipative nature of the problem we use a master equation for the density matrix of the system $\dot{\rho} = -i[H_{\rm RH},\rho]+\sum_n\mathcal{D}_n[\rho]$ where the Liouvillian has the form
\begin{align}
  \label{eq:3}
  \mathcal{D}_n[\rho] &= 
  \gamma \mathcal{L}[\hat \sigma_n^-,\rho]
  +
  \kappa \mathcal{L}[\hat a_n^{},\rho],
  \end{align}
with $\mathcal{L}[\hat X,\rho]=2 \hat X \rho \hat X^\dagger - \hat X^\dagger \hat X \rho - \rho \hat X^\dagger \hat X$ the Lindblad superoperator. Here $\kappa$ and $\gamma$ are (constant) decay rates. 

\begin{figure}[htpb]
  \centering
  \includegraphics[width=3.2in]{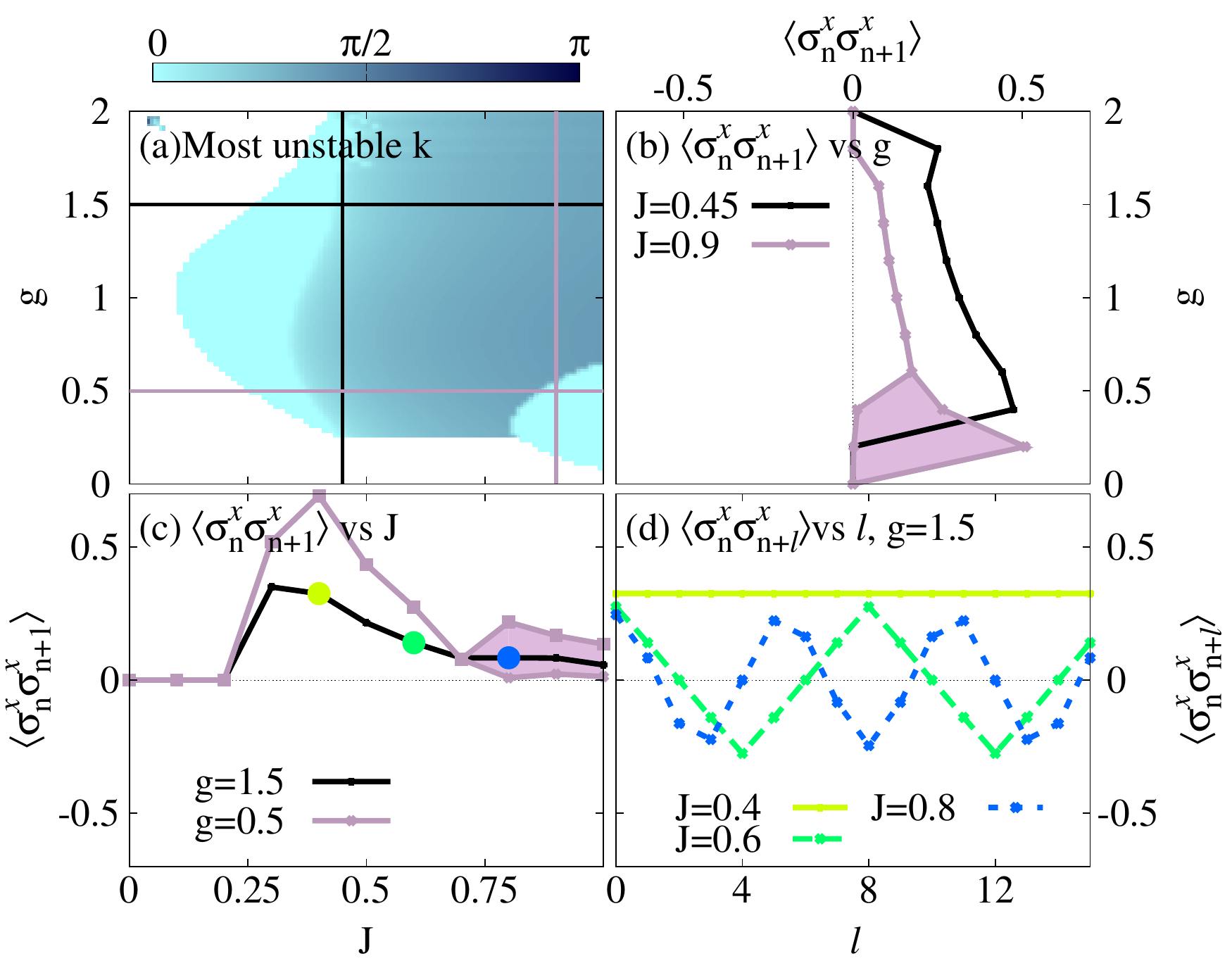}
  \caption{(a) Mean-field phase diagram of driven dissipative
    Rabi-Hubbard model at $g=g^\prime, \omega=\omega_0$, calculated by linear stability
    analysis of the normal state.  Color scale indicates the
    wavevector of the most unstable mode.  This wavevector predicts
    the ordering seen by finding the steady state solution for a chain
    of 16 cavities, as shown in panels (b-d).  Panels (b,c) show the
    nearest-neighbor correlations on the vertical and horizontal cuts
    marked in panel (a). The shaded region shows the envelope of the limit cycle oscillations 
    of the correlation function. Panel (d) shows the correlation vs
    separation at the three points marked in panel (c), revealing the
    incommensurate ordering. Parameters (also for the other figures): $\omega=\omega_0=1$,$\kappa=0.1$, 
    $\gamma=0.05$. }
  \label{fig:basic-pd}
\end{figure}

The RH Hamiltonian in Eq.~(\ref{eq:1}), as well as the dissipator in Eq.~(\ref{eq:3}), have a global $Z_2$ \emph{parity} symmetry, corresponding to a simultaneous change of sign of cavity and qubit operators, $\left(\hat a^\dagger,\hat\sigma^+\right)\rightarrow \left(-\hat a^\dagger,-\hat\sigma^+\right)$.  As a result, on general grounds we can expect a steady state phase diagram with a symmetric phase, where any quantity which is odd under parity will vanish, i.e. $\langle \hat\sigma^x_n\rangle = \langle \hat a^{}_n+\hat a_n^\dagger \rangle = 0$, and a phase with broken $Z_2$ parity symmetry.  

\emph{Mean Field Theory of Open Rabi Hubbard - } To characterize the steady state properties of open RH model
we make a mean-field ansatz for the system density matrix $\rho = \bigotimes_n \rho_n$. The dynamics reduces to a collection of  inequivalent single-site RH problems $\partial_t \rho_n =-i[\hat h_n,\rho_n]+\mathcal{D}_n[\rho_n] + i J 
  [\alpha^\ast_n \hat a^{}_n + \alpha^{}_n \hat a^\dagger_n, \rho_n]$ in a self-consistent field $\alpha_n = \sum_{m:\langle mn \rangle} \text{Tr}(\rho_m \hat a _m)$. Such an ansatz follows the standard concept of mean-field theory, that
each site sees only the average field of its neighbors~\cite{kadanoff2000statistical}.  
Thus, as for all mean-field theories, it becomes increasingly accurate in higher dimensions, as high coordination suppresses fluctuation contributions beyond the mean-field.

We start our discussion from the $g=g'$ case. In order to identify the phase boundary and to guide our analysis of the ordered phase, it is useful to first study the instability of the homogeneous normal state, by adding a small perturbation to the factorized density matrix as done in Ref.~\onlinecite{Boite2014}, i.e. $\rho = \bigotimes_n (\rho_{ss} + \delta \rho_n)$ where $\rho_{ss}$ is normal state density matrix obtained from the equation $\mathcal{M}_n[\rho]\equiv-i[\hat h_n,\rho_{ss}]+\mathcal{D}_n[\rho_{ss}]=0$ . Considering the one-dimensional case for simplicity and taking the fluctuations as plane waves of the form $\delta \rho_{n} =\sum_{k} \delta \rho_{k} e^{i (kn - \nu_{k} t)} + \text{H.c.}$ we obtain the equation of motion:
\begin{equation}
  \label{eq:4}
  -i \nu_{k} \delta \rho_{k} = \mathcal{M}_n[\delta \rho_{k}]
  - t_{k}  \left\{
    \text{Tr}(\hat{a} \delta \rho_{k}) i[\hat{a}^\dagger, \rho_{ss}] 
    + \text{H.c.} \right\},
\end{equation}
where  $t_{k}=-2J \cos(k)$ is the one-dimensional bare photon dispersion. A positive imaginary part of the frequency, $\mbox{Im}[\nu_k]>0$, signals the growth of fluctuations with momentum $k$ and the onset of normal state instability.
The results of linear stability analysis are plotted in Fig.~ \ref{fig:basic-pd}(a) where we can see the phase boundary in the $(g,J)$ plane and, in color scale, the wavevector of the most unstable mode. Two remarkable features immediately appear. First, the boundary has  a ``nose'' at small $J$, i.e.  a minimal critical $J$ required to enter the ordered phase.  This is in contrast to the ground state phase diagram~\cite{schiro_phase_2012}, in which the critical value of $J$
asymptotically approaches $0$ as $g= g^\prime \to \infty$. In addition, the nature of the broken symmetry phase itself shows an interesting evolution across the phase diagram. As the most unstable wavevector evolves smoothly from $k=0$, characteristic of a uniform ferroelectric (F) phase, toward $k=\pi/2$, the wavelength must pass through irrational values, corresponding to an instability towards incommensurate order. Such symmetry-broken inhomogeneous states requires to model a finite length array  and we consider in panels (b)-(d) a 16 site array with periodic boundary conditions.
We focus on correlations $\langle \hat\sigma^x_n \hat\sigma^x_{n+l}\rangle$ which at short distance ($l=1$, panels b-c) are ferroelectric but alternate in sign as a function of distance $l$ (panel d) revealing the inhomogeneous ordering. The finite length of the array is enough to see the trend of density-wave period vs hopping $J$,  although it prevents a continuous evolution of the period. Such incommensurate order is absent in equilibrium, where the minimum free energy state always has a constant phase across the array. Another unique feature of the steady state phase diagram
is the existence of limit cycles~\cite{Lee2011a,JinEtAlPRL13,LudwigMarquardtPRL13,Chan2014}. Within the linear stability analysis, a limit cycle can be anticipated if the normal state becomes unstable via a Hopf bifurcation~\cite{strogatz94} --- i.e. if there are a pair of eigenvalues $\nu_k = \pm \nu_k^\prime + i \nu_k^{\prime\prime}$ that simultaneously become unstable, leading to an oscillatory instability.  This in fact occurs for the region around $g=0.25, J>0.8$ where the most unstable $k$ returns to $k=0$.  The existence of the limit cycle is confirmed by direct time evolution of the equations of motion; in Fig.~\ref{fig:basic-pd}(b,c) the shaded region shows the envelope of the limit cycle oscillations of the correlation function. 

As we move away from the pure RH limit and consider $g\neq g'$ two main features arise~\cite{Note1}, namely (i) the shape of the phase boundary changes with multiple separate ordered regions and quite remarkably (ii) for certain values of  light-matter couplings $g',g$ we find an instability at $k=\pi$ corresponding to antiferroelectric (AF) order, where qubit and photon polarization alternates in sign between even and odd sites of the array, i.e. $\langle \sigma^x_n\rangle, \langle a_n+a^{\dagger}_n\rangle \sim (-1)^n$. This is a particularly striking result, considering that the effective qubit-qubit interaction in the equilibrium groundstate would be ferromagnetic~\cite{schiro_phase_2012}, and further pinpoints the profound differences between the open driven-dissipative, and equilibrium incarnations of the RH model. 

\begin{figure}[htpb]
  \centering
  \includegraphics[width=3.2in]{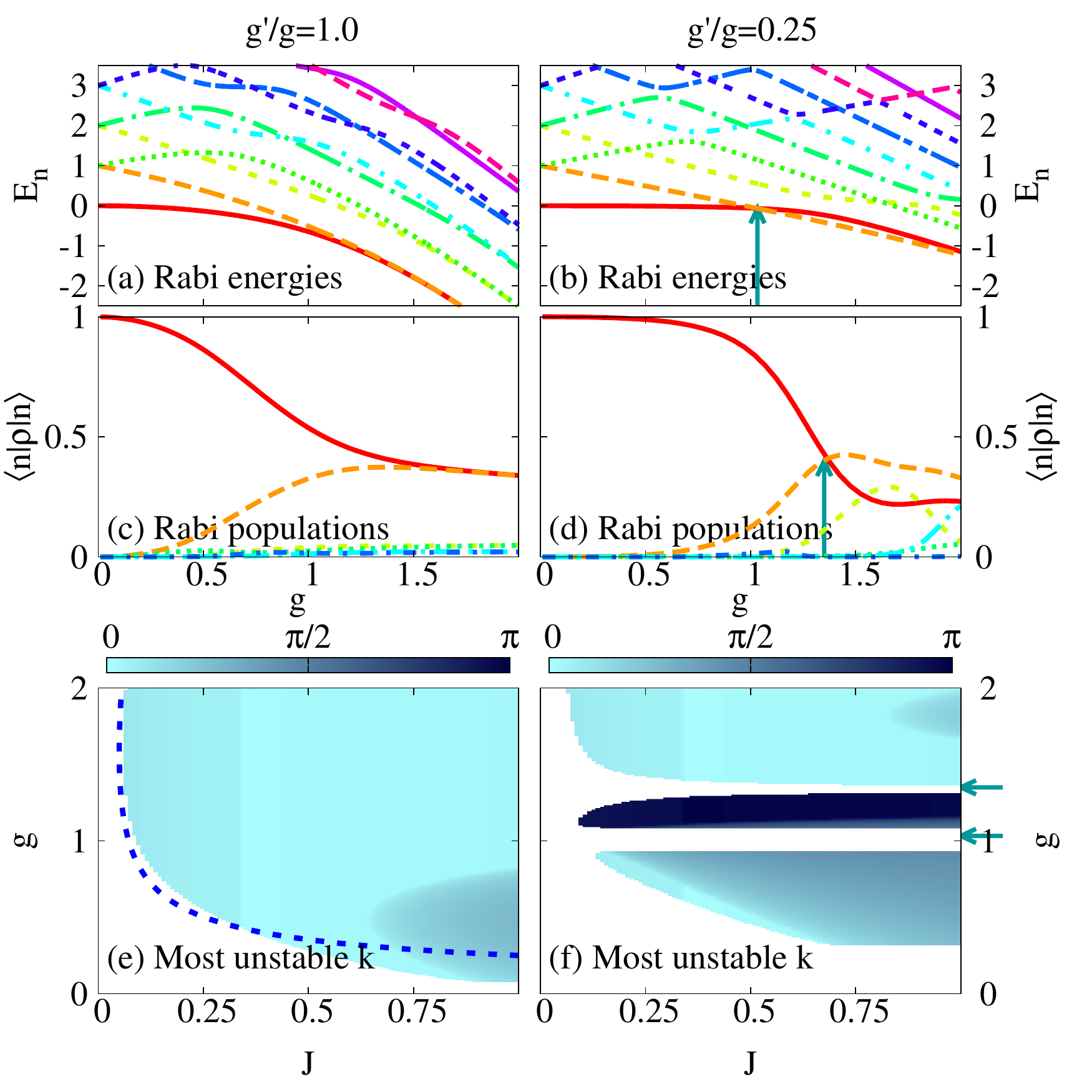}
  \caption{Properties of the effective spin model $g^\prime/g=1$
    (left) and $g^\prime/g=0.25$ (right).  (a,b) Eigenvalues and (c,d)
    normal state populations of the eigenstates of the Rabi model.
    The effective spin $1/2$ model truncates to only the first two levels:
    the solid (red) and long-dashed (orange) curves.  For
    $g^\prime/g=0.25$, the energies and populations of these levels
    cross, as marked by an arrow.  (e,f) Phase diagram of the
    effective Ising model as obtained by linear stability analysis, with
   color scale indicating the  wavevector of the most unstable mode, and by mean field analysis (dashed line).   
   Arrows in panel (f) mark the crossing points marked in panels (b,d).}
  \label{fig:effective-model}
\end{figure}
Even extending thermodynamics to 
negative effective temperature, it is not possible to explain all the features
identified, such as limit cycles or incommensurate order. Predicting the pattern of steady states clearly requires going beyond equilibrium thermodynamics.

\emph{Effective Spin Model - }  We now introduce an effective spin $1/2$ model which captures the essential physics of the RH model~\cite{Note1}. We start by considering the single site RH model, i.e. we set $J=0$ in Eq.~(\ref{eq:1}), and plot in figure Fig.~\ref{fig:effective-model} (a-d) the energies and the steady state populations of its eigenstates as a function of $g$,  for two different values of $g'/g$. We first consider the $g'=g$ case, panels (a,c). Here we notice that (i) the two low-lying states become almost degenerate for large $g$ and (ii) they are the only states effectively populated. The idea is then to truncate the local Hilbert space to the two lowest energy states of the on-site RH Hamiltonian that we denote $|{\pm}\rangle_n$ according to their (opposite) parity.  
Within this space the on-site Hamiltonian simply becomes $\hat h_n^{\text{eff}}=\Delta \, \hat{\tau}^z_n$ while the Liouvillian becomes
$\mathcal{D}^{\text{eff}}_n[\rho] = 
\gamma \mathcal{L}[S_-\hat{\tau}_n^- + S_+ \hat{\tau}_n^+, \rho] +
\kappa \mathcal{L}[A_-\hat{\tau}_n^- + A_+ \hat{\tau}_n^+, \rho]$,
where $\hat{\tau}^{i=x,y,z}_n$ are Pauli operators, and
$\Delta=E_{-}-E_{+}$ is the splitting between the lowest energy odd
and even parity states, and $A_\pm, S_\pm$ are the matrix elements
$A_\pm = {}_{n}\langle {\pm} | \hat a_n |{\mp}\rangle_{n}, S_{\pm} = {}_{n}\langle {\pm} |\hat \sigma^-_{n}|{\mp}\rangle_{n}$.
Note that the value of $\Delta,
A_\pm, S_\pm$ are all functions of $g, g^\prime, \omega, \omega_0$,
found by diagonalizing the Rabi model~\cite{schiro_phase_2012,Note1}. 
In addition, these local effective two-level systems are coupled by an anisotropic exchange mediated by photons, 
$J_{x,y}\sim J(A_-\pm A_+)$ which gives rise to an effective Hamiltonian in the transverse field Ising universality class.

We now show that the effective model captures the salient features of the RH model.
Firstly, in the limit of large $g=g^\prime$, one can show that there is an exponentially small splitting $\Delta=\omega_0 \exp(-2g^2/\omega^2)$ and the matrix elements become almost identical $A_{\pm}=(-1\pm\Delta/\omega)g/\omega$. As a consequence the hopping is predominantly an Ising coupling $\hat{\tau}^x_n \hat{\tau}^x_{n+1}$, and, in $d$ dimensions, one can derive a simple expression for the critical  hopping
\begin{equation}
  \label{eq:5}
  J_{\text{crit}} 
  \simeq
  \frac{1}{d}\left[
      \frac{\kappa^2 g^2}{\omega^3} + \frac{\omega^3}{16 g^2}\right].
\end{equation}
 Such an expression clearly explains the appearance of a minimum  
$J_\text{crit} > \kappa/2d$ for any finite loss, $\kappa$, as opposed to the exponentially small critical coupling $J_{\text{crit}}^{\text{eq}}\sim \Delta$ found in equilibrium. Furthermore Eq.~(\ref{eq:3}) matches
the linear-stability phase boundary remarkably well, see Fig.~\ref{fig:effective-model}(e).  In addition, it shows that as the loss $\kappa \to 0$, the nose will move toward $g\to \infty, J\to 0$ consistent with the equilibrium phase diagram~\cite{schiro_phase_2012}.

We next consider the case $g^\prime\neq g$, and plot in Fig.~\ref{fig:effective-model} (b,d), the energy levels and steady state population of the single site RH model. In this case there are energy levels crossings (see arrow), corresponding to a change in sign of local transverse field $\Delta= E_{-}-E_{+}$ for our effective spin model. 
This has interesting consequences for the lattice model as we see in panel (f) which shows the phase diagram as obtained from linear stability analysis for $g^\prime=g/4$. At the degeneracy point the ordered phase is suppressed, and beyond the crossing point an AF instability occurs (see bottom arrow), as recently observed in the transverse field Ising model~\cite{Joshi2013a}. Upon further increasing the coupling $g$ a further transition to a normal phase occurs, followed by a recovery of ferroelectric order (top arrow). This latter effect is associated with a population inversion between the $|{\pm}\rangle$ eigenstates, as one can see in the level occupations shown in Fig.~\ref{fig:effective-model}(d).  
 
 Thus, the sequence of F-AF-F can be explained as follows: At small $g$ the ground state
is that of even parity, and this state is the most occupied, leading
to F.  On increasing $g$, first the energy ordering of the odd and
even parity states is swapped, leading to AF, where the even parity
state is most occupied despite being of higher energy.  
Then, the occupation of the even and odd parity states inverts, so that once
again the lowest energy state is maximally occupied, and F ordering
is restored. 
\begin{figure}[!htb]
  \centering
  \includegraphics[width=3.2in]{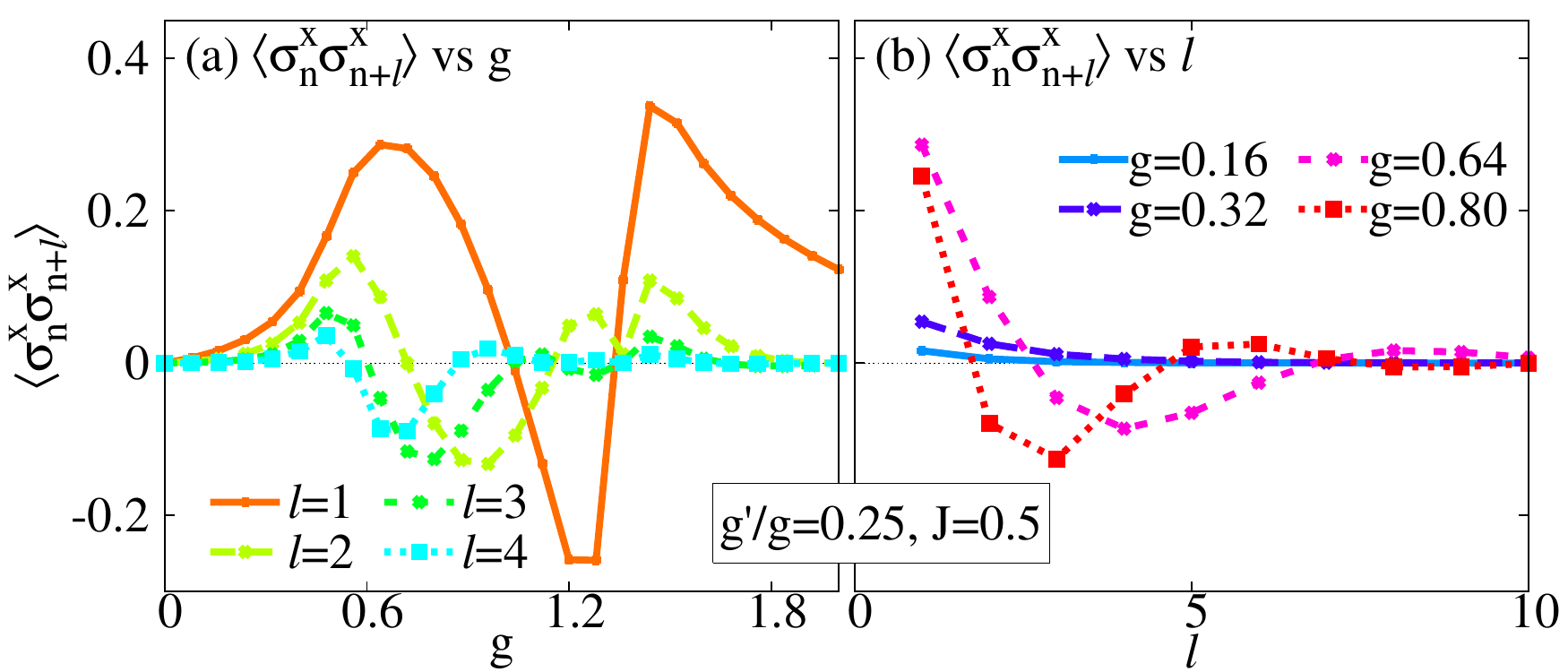}
  \caption{Correlations of the one-dimensional effective spin $1/2$
    model, evaluated in an infinite-MPO approach for $g^\prime/g=0.25,
    J=0.5$ (a) Correlations vs $g$, at various separations $l$ between
    sites.  (b) Correlations vs separation at various values of $g$.
    These confirm the ordering seen in mean-field theory, specifically
    the sequence of F-AF-F on increasing g, but show only
    short-range order as expected in one dimension.
   }
  \label{fig:mpo_vary-ratio}
\end{figure}
The qualitative picture emerging from the effective spin model is able to reproduce the salient features of the RH model,
both in terms of the phase boundary and in terms of pattern of broken symmetry phases~\cite{Note1}, at least for moderate values of $g$.  At yet higher coupling $g$, even higher states become occupied sequentially when resonances between excited state energy levels occur. The occupation of these higher levels  demonstrates the eventual failure of the effective spin $1/2$ model at large $g$. 

\emph{MPO Results - } A natural question is whether our mean field analysis survives to strong quantum fluctuations in low dimensions. In this respect the effective spin model has the advantage of being amenable to an exact numerical treatment with an infinite matrix product operator approach (iMPO), which we now turn to describe. 
 In Fig.~\ref{fig:mpo_vary-ratio} we show iMPO results for spin correlators evaluated for $g^\prime/g=0.25,  J=0.5$ as a function of $g$ at various separations $l$ between sites (panel a) and as a function of separation $l$ at various values of $g$. These numerically exact results confirm the ordering seen in mean-field,  specifically
    the sequence of F-AF-F upon increasing $g$, but additionally show that in low dimensions, fluctuations destroy long-range order in driven dissipative systems, as expected.  Further results are presented in the supplemental material~\cite{Note1} supporting this statement.

In summary, we have presented the steady-state phase diagram of the non-equilibrium Rabi-Hubbard model, using various mean-field-based techniques and a matrix product operator approach that can capture effects beyond mean-field. The phase diagram of the non-equilibrium model was found to be far richer than the equilibrium analogue, exhibiting ferroelectric, antiferroelectric and incommensurate ordering. In addition, the phase diagram was found to also exhibit limit-cycle solutions. The MPO results confirm qualitatively the pattern of the phases.

The research data supporting this publication can be
accessed  at  http://dx.doi.org/10.17630/a8829c09-7d49-
4472-8e9e-f8219b53d640. HET acknowledges support through NSF Grant
  No. DMR-1151810 and The Eric and Wendy Schmidt Transformative
  Technology Fund. CJ \& JMJK acknowledge support from the EPSRC
  program ``TOPNES'' (EP/I031014/1). JMJK acknowledges support from
  the Leverhulme Trust (IAF-2014-025). RF acknowledges support 
  from EU  (IP-SIQS and QUIC). We are grateful to C. Aron for help in preparing
  Fig. 1.  This work was supported in part by the National Science
Foundation under Grant No. NSF PHY11-25915.
 Author contributions: Mean field calculations were performed by MS, CJ and MB. MPO calculations
  by CJ.  RF, JK and HET supervised the work.   


%

\end{document}


\title{Supplementary Material for ''Exotic attractors of the non-equilibrium Rabi-Hubbard model''}
\author{M. Schir\'o}
\affiliation{Institut de Physique Th\'{e}orique, Universit\'{e} Paris Saclay, CNRS, CEA, F-91191 Gif-sur-Yvette, France}
\author{C. Joshi}
\affiliation{Department of Physics, University of York, Heslington, York, YO10 5DD, UK}
\affiliation{SUPA, School of Physics and Astronomy, University of St Andrews, St Andrews KY16 9SS, UK}
\author{M. Bordyuh}
\affiliation{Department of Electrical Engineering, Princeton University, Princeton, New Jersey 08544, USA}
\author{R. Fazio}
\affiliation{NEST, Scuola Normale Superiore and Istituto Nanoscienze-CNR, I-56127 Pisa, Italy}
\author{J. Keeling}
\affiliation{SUPA, School of Physics and Astronomy, University of St Andrews, St Andrews KY16 9SS, UK}
\author{H. E. T\"ureci}
\affiliation{Department of Electrical Engineering, Princeton University, Princeton, New Jersey 08544, USA}
\date{\today}

\maketitle

The supplementary material is organized as follows: Section~\ref{sec:raman-driving-scheme} presents further details of the possible explicit
driving scheme realizing a (tunable) Rabi-Hubbard model as presented
in Eq.~(1) of the main text.  Section~\ref{sec:meanfield_generalizedRH} presents the mean field and linear stability results for the generalized Rabi-Hubbard, with $g'\neq g$ that were mentioned in the main text.
Section~\ref{sec:effective-spin-model} discusses the construction of the effective spin model introduced in
the main text and the linear stability analysis of this model.
Section~\ref{sec:mpo-results-rabi} presents further results on the
effective spin model obtained by MPO simulations of an infinite chain.

\section{Raman Driving Scheme and Tunable Rabi Hubbard}
\label{sec:raman-driving-scheme}

In this section we present a possible scheme to obtain the Rabi Hubbard model starting from a driven four-level atom in a cavity. This scheme follows closely the original proposal~\cite{dimer_proposed_2007} for a driven realisation of the
generalized Dicke model, which was recently experimentally realized~\cite{baden_realization_2014}. 
To show most clearly how Rabi-like interactions can be generated from multi-atom driven problem we focus here on the isolated single cavity Hamiltonian, and disregard for now the photon hopping to other neighboring resonators. The interference between coupling to other cavities and pump induced terms can generate very weak long-ranged hopping and interactions between cavities. The analysis of these small corrections is left for future studies.

\begin{figure}[!htb]
  \centering
  \includegraphics[width=3.in]{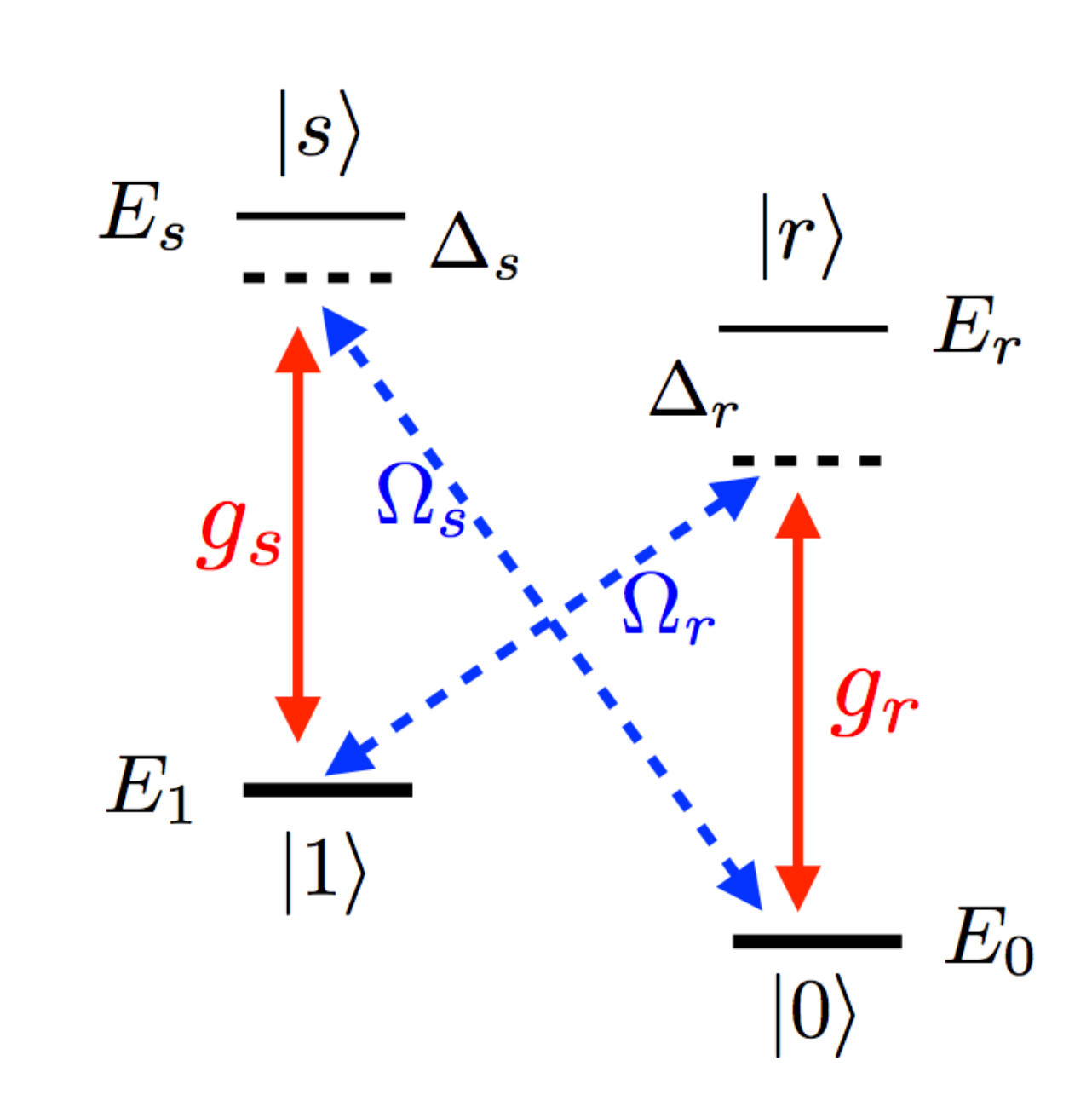}
  \caption{Cartoon of the level structure and Raman driving scheme to engineer a tunable Rabi Hamiltonian.}
  \label{fig:scheme}
\end{figure}

Our starting point is thus a four-level (artificial) atom in an optical cavity (or microwave resonator), supporting a single photon mode, see Figure~\ref{fig:scheme}. The full local Hamiltonian is made of several terms. The non-interacting atom and cavity are described by
\begin{displaymath}
\hat{H}_{0}=\omega_{\text{cav}}\,\hat{a}^{\dagger}\hat{a}+\sum_{j=0,1,r,s}\,E_j\,\vert\,j\rangle\langle\,j\vert.
\end{displaymath}
Following Ref.~\onlinecite{dimer_proposed_2007} we consider a case where the cavity mode drives transitions between the states $\vert0\rangle\leftrightarrow\vert\,r\rangle$ and $\vert1\rangle\leftrightarrow\vert\,s\rangle$. Selecting
these transitions can either be done by tuning the other transitions to be far off resonance with the cavity mode, or ideally, by engineering a system with selection rules restricting which transitions the cavity mode polarisation couples to~\cite{baden_realization_2014}. The resulting cavity-atom coupling Hamiltonian reads:
\begin{displaymath}
\hat{H}_{\text{int}}=g_r\,\vert\,r\rangle\langle\,0\vert\,\hat{a}
+
g_s\,\vert\,s\rangle\langle\,1\vert\,\hat{a}+ \text{H.c.}
\end{displaymath}

Since the goal is to generate effective light-matter interactions between the photon and the low-lying doublet $\vert\,0\rangle,\vert\,1\rangle$, we need to connect these states through resonant two-photon transitions. This can be done by adding the drive term:
\begin{displaymath}
\hat{H}_{\text{drive}}(t)= \frac{\Omega_r}{2}\,e^{-i\omega_r^p\,t}\,\vert\,r\rangle\langle\,1\vert +
\frac{\Omega_s}{2}\,e^{-i\omega_s^p\,t}\,\vert\,s\rangle\langle\,0\vert+\text{H.c.}
\end{displaymath}
A cartoon of the level scheme, and the pump- and cavity-mediated transitions
is shown in Figure~\ref{fig:scheme}.

The crucial feature of this scheme is that the laser drives transitions between $\vert1\rangle\leftrightarrow\vert\,r\rangle$ and $\vert0\rangle\leftrightarrow\vert\,s\rangle$ so that the combined effect of $H_{\text{int}}$ and $H_{\text{drive}}$ is to induce two-photon   transitions between $\vert\,0\rangle$ and $\vert\,1\rangle$, involving emission or absorption of a cavity photon. These will become the familiar rotating and counter-rotating terms of the Rabi model.

We proceed in two steps, that we briefly outline here. 
First, we eliminate the explicit time dependence of the drive by performing an unitary transformation to a rotating (comoving) frame. This can be implemented by the operator:
\begin{displaymath}
  \hat{\Omega}(t)=\exp\left(-i\hat{K}t\right), \quad
  \hat{K}=\alpha\,\hat{a}^{\dagger}\hat{a}+\sum_i\,\beta_i\vert\,i\rangle\langle\,i\vert,
\end{displaymath}
which has the property that, $\hat{\Omega}^{\dagger}(t)\,\hat{a}\, \hat{\Omega}(t) = \hat{a} e^{-i\alpha t}$ and $\hat{\Omega}^{\dagger}(t)\,
\vert\,i\rangle\langle\,j\vert\, \hat{\Omega}(t) = 
\vert\,i\rangle\langle\,j\vert\,e^{i\left(\beta_i-\beta_j\right)t}$. We can fix the parameters in $\hat{K}$ in such a way that the transformed Hamiltonian 
$\tilde{H}=i\partial_t\left(\hat{\Omega}^{\dagger}(t)\right)\hat{\Omega}(t)+\hat{\Omega}^{\dagger}(t)\,\hat{H}(t)\,\hat{\Omega}(t)$, with $\hat{H}=\hat{H}_0+\hat{H}_{\text{int}}+\hat{H}_{\text{drive}}$, becomes  time-independent.
We stress that while the problem in this new rotating frame is now fully time-independent, it remains intrinsically out of equilibrium in nature since (i) we are forced to study the dynamics of $\tilde{H}$, and (ii) the distribution function of any bath modes the system couples to becomes strongly non-thermal (breaking detailed balance between gain and loss), as the unitary transform $\hat{\Omega}$ must also be applied to the system-bath coupling.  

We now proceed to the elimination of atomic excited states $\vert\,r\rangle,\vert\,s\rangle$ in order to obtain an Hamiltonian for the manifold $\vert\,i\rangle=\vert\,0\rangle,\vert\,1	\rangle$ that will play the role of our qubit. This can be done perturbatively, using a Schrieffer-Wolf transformation, 
$\tilde{H}_{\text{S}}=e^{i\hat{S}}\tilde{H}e^{-i\hat{S}}$ with a generator $\hat{S}$ coupling the qubit manifold  $\vert 0\rangle,\vert 1\rangle $ to the excited states $\vert r\rangle,\vert s\rangle$
\begin{multline*}
\hat{S}=
i
\biggl(\frac{\Omega_r}{2 \Delta_r} \vert r\rangle \langle 1\vert
    + \frac{\Omega_s}{2 \Delta_s} \vert s\rangle \langle 0\vert+\\
    + \frac{g_r}{\Delta^\omega_r}\,\hat{a}\vert r\rangle \langle 0\vert
    + \frac{g_s}{\Delta^\omega_s}\,\hat{a}\vert s\rangle \langle 1\vert
    \biggr) + H.c..
\end{multline*}
This is obtained by imposing the condition $[\hat{H}_0,i\hat{S}]=\hat{H}_{\text{int}}+\hat{H}_{\text{drive}}$ (in terms of the time-independent Hamiltonian).
We have introduced the detunings 
\begin{align*}
\Delta^{\omega}_{r}
&=\omega_{\text{cav}}+E_0-E_r,\\
\Delta^{\omega}_{s}
&=\omega_{\text{cav}}+E_1-E_s+(\omega^s_p-\omega_p^r)/2,\\
\Delta_r &= E_1-E_r+\omega_p^r,\qquad
\Delta_s=E_0-E_s+\omega_p^s.
\end{align*}
To leading order in the strength of the drive and light-matter coupling we get 
$\tilde{H}_{\text{S}} = \tilde{H}_{\text{Rabi}} + \mathcal{O}(g_r/\Delta_r, g_s/\Delta_s, \Omega_r/\Delta_r, \Omega_s/\Delta_s)$ where
\begin{multline}
  \label{eq:2}
  \tilde{H}_{\text{Rabi}}= \omega\,\hat{a}^{\dagger}\hat{a}
  +\frac{\omega_0}{2}\hat{\sigma}^z+\lambda\,\hat{a}^{\dagger}\hat{a}\hat{\sigma}^z+
  g (\hat{a}^{\dagger}\hat{\sigma}^- + \hat{a}^{}\hat{\sigma}^+)
  \\+
  g^\prime (\hat{a}^{\dagger}\hat{\sigma}^+ + \hat{a}^{}\hat{\sigma}^-).
\end{multline}
This is a {\it generalized Rabi model}, for which the parameters depend on the frequency and strength of the external drive as follows:
\begin{align*}
\omega&=\omega_{cav}+ \frac{1}{2}\left(\frac{g^2_r}{\Delta^{\omega}_r}+\frac{g^2_s}{\Delta^{\omega}_s}\right)-\frac{\omega^s_p+\omega^r_p}{2}, \\
\omega_0 &=   E_1-E_0  +\frac{1}{8}\left(\frac{\Omega^2_r}{\Delta_r}-\frac{\Omega^2_s}{\Delta_s}\right)
-\frac{\omega^s_p-\omega^r_p}{2},\\
\lambda &= \frac{1}{2}\left(\frac{g^2_r}{\Delta^{\omega}_r}-\frac{g^2_s}{\Delta^{\omega}_s}\right),\quad
g = \frac{g_r\Omega_r}{2\bar{\Delta}_r}, \quad
g^\prime = \frac{g_s\Omega_s}{2\bar{\Delta}_s},
\end{align*}
where we defined $1/\bar{\Delta}_{r/s} = 1/\Delta^{\omega}_{r/s} + 1/\Delta_{r/s}$. 

To complete the mapping to the generalized Rabi model we must choose parameters
such that the coefficient $\lambda=0$, as also discussed in~\cite{dimer_proposed_2007}.  To this end we note that $\Delta^{\omega}_r$ is drive-independent and therefore fixed by the physical realization, while $\Delta^{\omega}_s$ can be tuned by $\omega^s_p$. We therefore impose the condition:
\begin{displaymath}
\Delta^{\omega}_s = \Delta^{\omega}_r\frac{g_s^2}{g_r^2}, \qquad
\end{displaymath}
which immediately gives $\lambda=0$. From this we see that we may want to have $g_s/g_r\sim 1$ in order to have both $\Delta^{\omega}_r,\Delta^{\omega}_s$ large. The second drive frequency $\omega^r_p$ can then be used to control the effective detuning between the qubit and the resonator. In addition to changing the detuning through the pump frequencies, we can
vary the pump strengths $\Omega_r,\Omega_s$ to tune (independently) the relative strengths of the rotating and counter-rotating terms.

\section{Open Rabi Hubbard for $g'\neq g$}\label{sec:meanfield_generalizedRH}

\begin{figure}[t]
  \centering
  \includegraphics[width=3.2in]{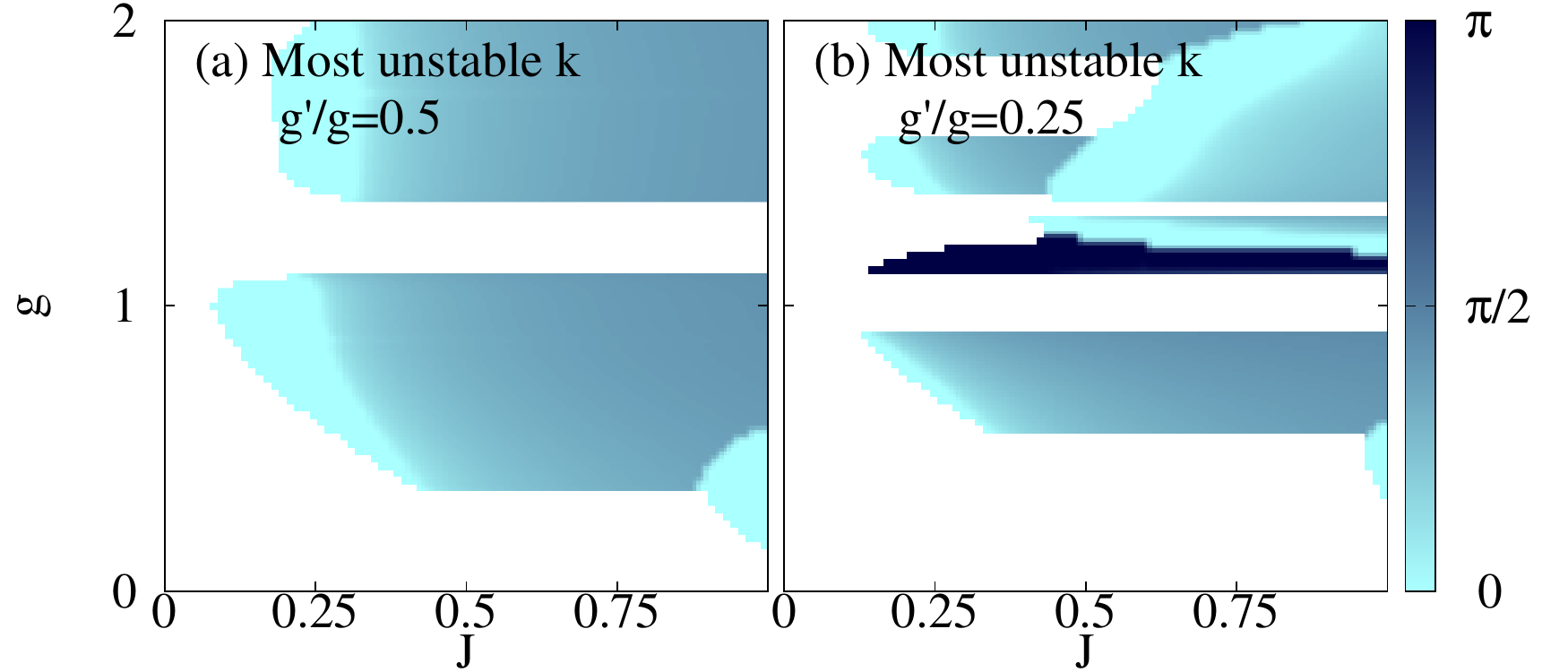}
  \caption{Mean field phase diagram of the generalized ($g'\neq g$) open Rabi Hubbard model as obtained from the linear stability analysis of the normal phase. We plot the phase boundary in the $(g,J)$ plane and, in color scale, the wavevector of the most unstable mode, for two different values of $g'/g$, respectively $g'=0.5 g$ (left panel) and $g'=0.25 g$ (right panel).}
  \label{fig:vary-ratio}
\end{figure}

In this section we present results, anticipated in the main text, for the phase diagram of the generalized open Rabi Hubbard, with $g'\neq g$.  We use again linear stability analysis of the normal phase, as described in the main text, and plot in figure~\ref{fig:vary-ratio} the phase boundary in the $(g,J)$ plane and, in color scale, the wavevector of the most unstable mode, for two different values of $g'/g$, respectively $g'=0.5 g$ (left panel) and $g'=0.25 g$ (right panel).  These results show several interesting features. Firstly there is a change in the topography of the phase boundary, with multiple separate ordered regions, as opposed to the case $g'=g$ which features a single, connected, broken symmetry phase. As a result, upon increasing the light-matter interaction $g$ one can have a sequence of transitions, where the symmetry is broken first, then restored and then broken again. In addition, for smaller values of the ratio $g'/g$ the nature of the ordered phase changes qualitatively and an instability at $k=\pi$, eventually emerges.  This corresponds to an antiferroelectric (AF) order, where photon and two-level system (qubit) polarization changes sign between neighboring sites. This latter result is particularly striking if interpreted in terms of equilibrium physics. Indeed the effective qubit-qubit interaction mediated by a population of photon modes in equilibrium at zero temperature would be negative at short distance~\cite{schiro_phase_2012} leading to a uniform ferroelectric pattern in the ground state. The results of figure~\ref{fig:vary-ratio} show the profound differences between the open driven-dissipative and equilibrium incarnations of the Rabi Hubbard model. As we have discussed in the main text, these intriguing results can be understood qualitatively in terms of an effective open-dissipative spin model whose construction we are now going to discuss in detail in the rest of this supplementary material.

\section{Effective Spin Model}
\label{sec:effective-spin-model}
By truncating the on-site Rabi model ($\tilde{H}_{\text{Rabi}}$ obtained in the previous section) to the lowest two eigenstates and labelling these as $\vert\pm\rangle_{n}$ (with corresponding energies $E_{\pm}$)
to denote their different parities, we obtain the following low-frequency effective Master equation describing the dynamics of the generalized Rabi-Hubbard model:
\begin{multline}
  \label{eq:1}
\partial_t\rho = -i[\hat{H}^{\text{\text{eff}}},\rho]+\kappa \mathcal{L}[A_-\hat{\tau}_n^- + A_+ \hat{\tau}_n^+, \rho]+\\
+\gamma \mathcal{L}[S_-\hat{\tau}_n^- + S_+ \hat{\tau}_n^+, \rho],
\end{multline}
where the effective Hamiltonian for the generalized Rabi-Hubbard Model reads (for a detailed discussion of this projection, see Refs.~\onlinecite{schiro_phase_2012,schiro_quantum_2013} which discuss the case $g=g^\prime$)
\begin{multline*}
\hat{H}^{\text{\text{eff}}}=\frac{\Delta}{2}\sum_n \hat{\tau}^z_n \\ - J\sum_{\langle n,m\rangle}
\left(A_-\hat{\tau}^{+}_n+A_+\hat{\tau}^-_n\right)
\left(A_-\hat{\tau}^{-}_m+A_+\hat{\tau}^+_m\right),
\end{multline*} 
where $\hat{\tau}_n^i$ are Pauli matrices on the $n$th site, $\Delta=E_+-E_-$ is the lowest excitation energy of the local Rabi model, and we 
have introduced the \emph{real-valued} coefficients
\begin{equation}
A_{\pm}={}_{n}\langle\pm\vert \hat{a}_{n}\vert\mp\rangle_{n},\qquad
S_{\pm}={}_{n}\langle\pm\vert \hat{\sigma}^-_{n}\vert\mp\rangle_{n}.
\end{equation}
For later convenience it is useful to introduce the even and odd combinations of these coefficients, namely
\begin{equation}
\alpha_{\pm}=A_+\pm A_-,\;\qquad\;\varsigma_{\pm}=S_+\pm S_-.
\label{eqSMmatrixel}
\end{equation}
The effective-model parameters $\Delta, \alpha_\pm, \varsigma_{\pm}$ are
  functions of the parameters $g,g^\prime, \omega_0$ in the Rabi
  model.  We will discuss below analytic approximations in the limit
  of large $g=g^\prime$, the effective model is however useful more generally.

\subsection{Mean-field theory, linear stability}
\label{sec:mean-field-theory}

The effective model can be used to find a closed-form expression for the critical hopping required to stabilize  the ordered state within the mean-field theory, as discussed next. Within a mean-field decoupling of the hopping, the single site Hamiltonian has the form 
\begin{equation}
\hat{H}^{\text{eff,MF}}_n = \frac{\Delta}{2}\hat{\tau}^z_n-\frac{J}{4}\alpha_+^2\,\hat{\tau}^x_n\,h^x_n-\frac{J}{4}\alpha_-^2\,\hat{\tau}^y_n\,h^y_n,
\end{equation} 
where the local effective fields 
$$
h^x_n=\sum_{m:\langle mn \rangle} x_m,\,\qquad\,h^y_n=\sum_{m:\langle mn \rangle} y_m,,
$$
are written in terms of $x_n=\langle \hat{\tau}^x_n\rangle, y_n=\langle \hat{\tau}^y_n\rangle, z_n=\langle \hat{\tau}^z_n\rangle$,
account for the effect of the sites $m$ connected by photon hopping to sites $n$.
 By introducing the damping rates
$$
\Gamma_{\pm}=\kappa\alpha_{\pm}^2+\gamma\varsigma_{\pm}^2,
$$
and the normal state inversion
$$
z_0 = \frac{2}{\Gamma_++\Gamma_-}\left(\kappa\alpha_+\alpha_-+\gamma\varsigma_+\varsigma_-\right),
$$
we can write down the mean field dynamics in a compact form as 
\begin{align*}
\partial_t x_n &= -\Delta y_n -\Gamma_- x_n -\frac{J}{2}\alpha_-^2\,h^y_n\,z_n,\\
\partial_t y_n &= \Delta x_n -\Gamma_+ y_n+\frac{J}{2}\alpha_+^2\,h^x_n\,z_n,\\
\partial_t z_n &= \left(\Gamma_++\Gamma_-\right)\left(z_0-z_n\right) -\frac{J}{2}\alpha_-^2\,h^x_n\,y_n+\frac{J}{2}\alpha_-^2\,h^y_n\,x_n.
\end{align*}
The normal state solution is given by $z_n=z_0$ and $x_n=y_n=0$. Considering small fluctuations about this stationary solution we obtain a decoupled equation for $z_n$, with damping rate $\Gamma_++\Gamma_-$ while for the other components the ansatz 
$x_n = \sum_{\vec{k}}\,x_{\vec{k}}\,e^{i\left(\vec{k}\cdot\vec{n}-\nu_{\vec{k}} t\right)}$ gives the secular equation for the frequencies
$\nu_{\vec{k}}$, 
\newcommand{\z}{z_0}
\begin{displaymath}
  (i\nu_{\vec{k}} -\Gamma_-)(i\nu_{\vec{k}}-\Gamma_+) +
  \left(\Delta - t_\vec{k} \alpha_+^2 \z\right) 
  \left(\Delta - t_\vec{k} \alpha_-^2 \z\right)=0,
\end{displaymath}
where $t_{\vec{k}}=-2J \sum_{i=1}^d\cos(k_i)$ is the $d$-dimensional
bare photon dispersion.  The instability, corresponding to a pitchfork
bifurcation, is given by $\nu_{\vec{k}}=0$. This leads to a simple
  expression for the critical $J$ in the limit $\vert
  \alpha_-\vert \ll \vert\alpha_+\vert$, which, as we will see, is satisfied at large $g=g^\prime$. 
For the case where a ferromagnetic (anti-ferromagnetic) instability
${k}_i=0(\pi)$ occurs, this expression is:
\begin{equation}\label{eqn:Jcrit}
J_{crit}=\mp\frac{1}{2d\alpha_+^2z_0}\left(\Delta+\frac{\Gamma_+\Gamma_-}{\Delta}\right).
\end{equation}
Since we consider only positive $J$, it
is clear from the form of this equation that the sign of
$t_{\vec{k}}$ required, i.e. whether the instability is ferro (F)
or antiferro (AF), is determined by the sign of the product
$\Delta \z$.  For $g=g^\prime$, we always have
$\Delta>0, \z<0$ and so only the ferromagnetic case  (negative sign) occurs
Thus, as discussed in the main text, the level crossing
at $\Delta=0$, and the inversion point at $\z=0$ lead to suppression
of ordering, and a switch between F and AF ordering.

\subsection{Large $g$ limit and Effective Model parameters}

In the limit of large $g=g^\prime$ one may use an approximate solution of the on-site Rabi model
to derive simple analytic expressions for the lowest excitation energy $\Delta$ and the matrix elements $\alpha_{\pm},\varsigma_{\pm}$.
To extract these expressions we start from the single site Rabi Hamiltonian 
\begin{equation}
\hat{H}^{\text{Rabi}}=\frac{\omega_0}{2}\hat{\sigma}^z+\omega \hat{a}^{\dagger}\hat{a}+g(\hat{a}^{\dagger}+\hat{a})\hat{\sigma}^x,
\end{equation}
and perform the unitary transformation 
$$
\hat{U}=\exp\left[ -\hat{X}\hat{\sigma}^x\right],\,\qquad\, 
  \hat{X}=\frac{g}{\omega}(\hat{a}^{\dagger}-\hat{a}),
$$
to obtain $\tilde{H}^{\text{Rabi}}=\hat{U}^{\dagger}\hat{H}^{\text{Rabi}} \hat{U}$ in the form
\begin{equation}
\tilde{H}^{\text{Rabi}}=-\frac{g^2}{\omega}+\omega \hat{a}^{\dagger}\hat{a} + \frac{\omega_0}{2}
\left[\cosh 2\hat{X}\,\hat{\sigma}^z-i\sinh 2\hat{X}\,\hat{\sigma}^y\right].
\end{equation}
In absence of the term in brackets the spectrum has an infinite sequence of two-fold degenerate states with energies $E^0_n=-(g^2/\omega)+n\omega$ and corresponding eigenstates $\vert \sigma=\{ \uparrow,\downarrow \}, n\rangle$ in the transformed basis. In what follows we carry out a perturbation expansion in $\delta\hat{H} = \frac{\omega_0}{2} \left[\cosh 2\hat{X}\,\hat{\sigma}^z-i\sinh 2\hat{X}\,\hat{\sigma}^y\right]$ for the states in the lowest manifold (identified by $n=0$). This is justified by the smallness of the parameter $\langle 0\vert e^{\pm 2\hat{X}}\vert 0 \rangle=\exp\left(-2(g/\omega)^2\right)$ in the large-g limit $g/\omega \gg 1$. To lowest order, this splits the $n=0$ doublet and yields an analytic expression for the lowest excitation energy of the Rabi model:
 \begin{equation}
\Delta = \omega_0\exp\left(-2g^2/\omega^2\right).
\end{equation}
To obtain an analytic expression for the matrix elements Eq.~\ref{eqSMmatrixel}, we also need to compute the first order correction to the wavefunctions $\vert \sigma, 0 \rangle$:
\begin{equation}
\vert \widetilde{{\sigma}, 0} \rangle = \vert {\sigma},0\rangle -\sum_{n>0,\tau}\frac{\vert{\tau},n\rangle\langle \tau, n\vert\delta\hat{H}\vert{\sigma},0\rangle}{n\omega},
\end{equation} 
where $\vert\widetilde{\sigma,0} \rangle$ denotes the corrected wavefunction, and the matrix elements of the
perturbation Hamiltonian can be found using the following expression:
\begin{align}\label{eqn:On}
\langle  n\vert\delta\hat{H}\vert0\rangle
&=
 \frac{\omega_0}{2}\langle n\vert \cosh(2\hat{X})\hat{\sigma}_z-i\sinh(2\hat{X})\hat{\sigma}_y\vert0\rangle\nonumber\\
&=\frac{\Delta}{2\sqrt{n!}}\left(\frac{2g}{\omega}\right)^n\left[I_{n\in even}\,\hat{\sigma}_z-i\,I_{n\in odd}\,\hat{\sigma}_y\right].
\end{align}
where $I_{n\in A}$ is an indicator function, taking value $1$ for $n\in A$ and zero elsewhere.

We can now evaluate the matrix elements of photon and spin operators in the space spanned by the low-energy doublet,
that in the transformed basis are simply the states $\vert \uparrow,0\rangle, \vert\downarrow,0\rangle$. To evaluate the spin matrix elements we note that, under the unitary transformation we have
\begin{equation}
\hat{U}^{\dagger}\hat{\sigma}^-\hat{U}=\frac{1}{2}\left[\hat{\sigma}_x+\cosh(2\hat{X})(-i\hat{\sigma}_y)+\sinh(2\hat{X})\hat{\sigma}_z\right],
\end{equation}
from which we conclude that
\begin{align}
S_+ &= \langle \uparrow,0\vert \hat{U}^{\dagger}\hat{\sigma}^-\hat{U}\vert\downarrow,0\rangle = \frac{1}{2}\left(1-\frac{\Delta}{\omega_0}\right),\\
S_- &= \langle \downarrow,0\vert \hat{U}^{\dagger}\hat{\sigma}^-\hat{U}\vert\uparrow,0\rangle = \frac{1}{2}\left(1+\frac{\Delta}{\omega_0}\right),
\end{align}
where to leading order we have only accounted for the transformation of the operator. Turning to the photon matrix elements we first note that under the unitary transformation we have
\begin{equation}
\hat{U}^{\dagger}\hat{a}\hat{U} = \hat{a}-\frac{g}{\omega}\hat{\sigma}_x,
\end{equation}
therefore to evaluate the matrix element one has to use the perturbed eigenstate to obtain, at leading order,
\begin{equation}
A_{+}=\langle \widetilde{\uparrow , 0} \vert \hat a-\frac{g}{\omega}\hat{\sigma}_x\vert\widetilde{\downarrow , 0} \rangle=\frac{g\Delta}{\omega^2}-\frac{g}{\omega},
, \quad
A_{-}=-\frac{g\Delta}{\omega^2}-\frac{g}{\omega}.
\end{equation}
Using these results, we find that the matrix elements can be found to leading order to be $\alpha_+ = 2 g/\omega, \alpha_- = 2 g \Delta/\omega^2, \varsigma_+=1, \varsigma_-=\Delta/\omega_0$, from which we can obtain the approximate expression for the critical boundary given in the main text.

\section{MPO Results for the Rabi Hubbard case, $g=g^\prime$}
\label{sec:mpo-results-rabi}

In order to test the predictions of mean-field theory, we may use
  a infinite Matrix-Product Operator (iMPO) approach~\cite{VidalPRL03,VidalPRL04,ZwolakVidalPRL04,SchollwockAnnPhys11} to find the
  non-equilibrium steady state of the effective model,
  supplementary Eq.~(\ref{eq:1}).  The iMPO approach is applicable to a
  translationally invariant problem~\cite{OrusVidalPRB08}, allowing one to evolve only a
  representative pair of sites, and has no finite-size effects.
  Despite this, as seen in Fig.~5 of the paper, iMPO can fully
  describe inhomogeneous order; this is because multi-site
  expectations involve traces of products of the representative
  matrix.  Our approach, as in previous work~\cite{Joshi2013a}, is to simulate an adiabatic sweep, starting from the point $g = g^\prime = 0$ where the state is trivially a product state. Such a sweep avoids the high transient entanglement that
  occurs with a quench of parameters, and means that the correlation
  functions we measure are converged for the bond dimension $\chi=120$
  we use.

\begin{figure}[!htb]
  \centering
  \includegraphics[width=3.2in]{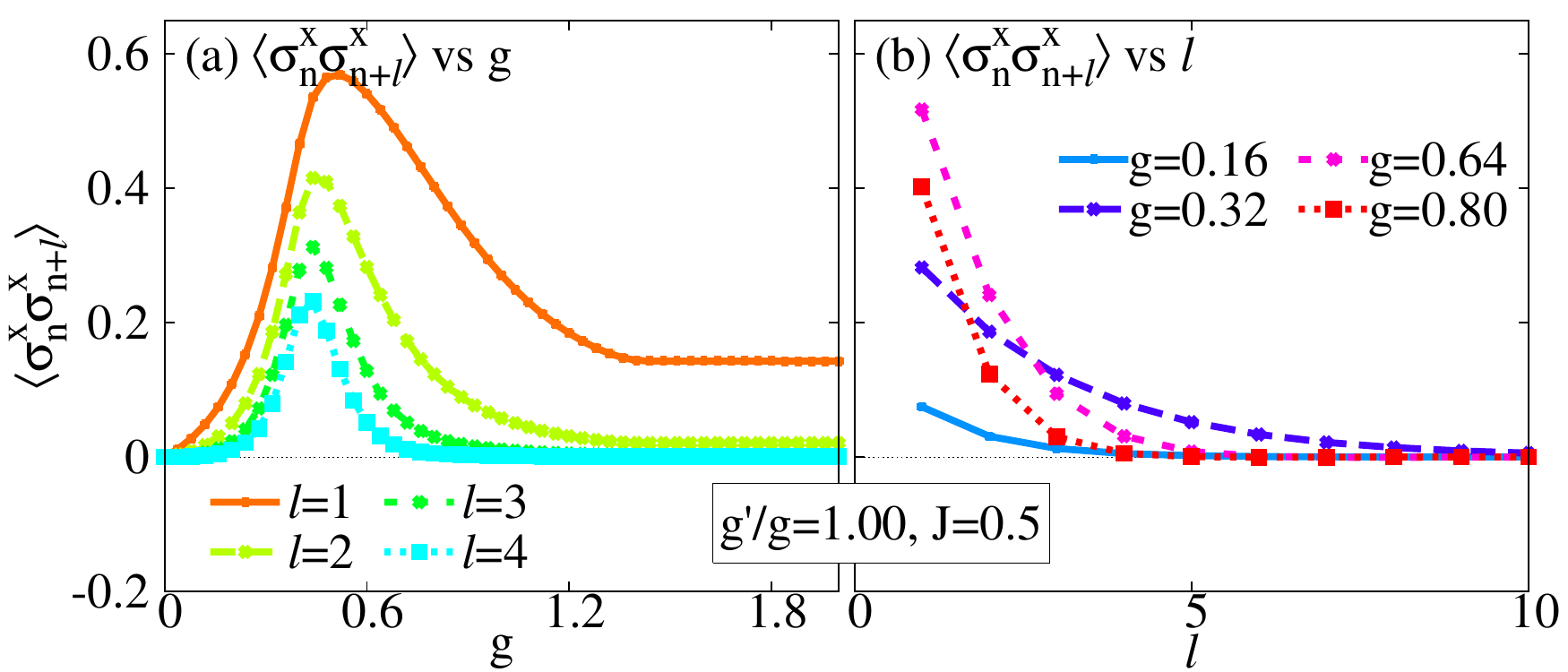}
  \caption{Correlations of the one-dimensional effective spin $1/2$
    model, evaluated in an infinte-MPO approach for $g^\prime=g=1, 
    J=0.5$ (a) Correlations vs $g$, at various separations $l$ between
    sites.  (b) Correlations vs separation at various values of $g$.
   }
  \label{fig:mpo_SM}
\end{figure}

In the main text we showed results for $g^\prime/g=0.25$, illustrating that the
F-AF-F sequence predicted by mean-field theory still exists (albeit short-ranged) for the MPO simulation.
  Here we present additional MPO results on the effective
  spin model in the pure Rabi case, i.e. for $g=g^\prime=1$. 
In particular the data reported in figure~\ref{fig:mpo_SM} show short-ranged ferroelectric correlations developing for several values of the light-matter interaction, consistently with the mean field picture. 

%